\documentclass[12pt]{article}

\usepackage{amsmath,amsfonts,amssymb,epsfig,psfrag,subfigure}
\usepackage{amsmath,amsfonts,amssymb}

\renewcommand{\ge}{\geqslant}

\newcommand{\imi}{\textrm{i}}
\renewcommand{\vec}[1]{\boldsymbol{#1}}

\begin{document}
\numberwithin{equation}{section}

\title{ Non-principal surface waves\\in deformed incompressible
materials}
\author{
  Michel Destrade, M\'elanie Ott\'enio, \\
  Alexey V.~Pichugin, Graham A.~Rogerson}
\date{2005}
\maketitle

\bigskip

\begin{abstract}
The Stroh formalism is applied to the analysis of infinitesimal
surface wave propagation in a statically, finitely and homogeneously
deformed isotropic half-space.
The free surface is assumed to coincide with one of the principal
planes of the primary strain, but a propagating surface wave is not
restricted to a principal direction.
A variant of Taziev's technique
[\textit{Sov. Phys. Acoust.} \textbf{35} (1989) 535]
is used to obtain an explicit expression of the secular equation for
the surface wave speed, which possesses no restrictions on the form
of the strain energy function.
Albeit powerful, this method does not produce a unique
solution and additional checks are necessary.
However, a class of materials is presented for which an exact secular
equation for the surface wave speed can be formulated.
This class includes the well-known Mooney-Rivlin model.
The main results are illustrated with several numerical examples.
\end{abstract}

\newpage

\section{Introduction}

The study of small-amplitude surface waves propagating in finitely
and homogeneously deformed hyperelastic materials has quite a long
history, dating back to the classical paper by Hayes and Rivlin
\cite{HaRi61} in 1961.
The interest in using the theory of small motions superimposed
on a large static deformation of a hyperelastic half-space is
manifold, for once the problem is solved the results are applicable
to various advanced topics. These, in particular, include the
non-destructive evaluation of solids (see Guz' and Makhort
\cite{GuMa00} for a review), the incremental stability analysis of
the loaded surface of a deformed material (see Guz' \cite{Guz99} for
comprehensive review and bibliography), and the acousto-elastic
effect (second-order theory of linear elasticity) where
the pre-deformation is also considered small (see Pao et al.
\cite{PaSF84} for a review of both experimental and theoretical
results).

The results of Hayes and Rivlin, valid for a compressible hyperelastic
material, have since been extended to the case of hyperelastic
materials subject to incompressibility \cite{DoOg90} or to a
generic isotropic internal constraint \cite{DeSc04}.
However, these works are limited to the consideration of
\textit{principal} surface waves, i.e. surface waves that propagate and
attenuate along principal directions of pre-strain.
We note that Connor and Ogden \cite{CoOg95} and Destrade
and Ogden \cite{DeOg04} considered two-partial (non-principal) surface
waves polarized in a principal plane of pre-strain.
Nevertheless, for \textit{three-partial} non-principal surface waves
very few explicit results exist and the scope of their
applicability is limited.
Specifically, Flavin \cite{Flav61} considered the problem for a
Mooney-Rivlin material with one material parameter much smaller than
the other;
Willson \cite{Will73,Will74} studied materials subject to equi-biaxial
pre-deformations;
Gerard \cite{Gera69}, Gerhart \cite{Gerh76}, and Iwashimizu and
Kobori \cite{IwKo78} worked with the linearized theory of second-order
acousto-elasticity; Chadwick and Jarvis
\cite{ChJa79}, Mase and Johnson \cite{MaJo87}, and
Chadwick \cite{Chad97} used the
Stroh-Barnett-Lothe integral formalism;
Rogerson and Sandiford \cite{RoSa99} used numerical methods for
an implicit secular equation; etc.

This paper presents an explicit secular equation for
the speed of a surface wave propagating in a principal plane, but not
in a principal direction, of a tri-axially deformed,
general hyperelastic incompressible material.
This result is achieved by using methods first developed by
Taziev \cite{Tazi87,Tazi89} for surface waves propagating in the
symmetry plane of a crystal.
Although similar, the analysis presented here reveals
some features particular to the context of nonlinear elasticity.
In general, the surface wave consists of a linear
combination of three partial modes, each proportional to
$\exp\left\{\imi k(\vec{n}\cdot\vec{x} + q_{\alpha} \vec{m}\cdot\vec{x} - vt)\right\}$,
$\alpha=1,2,3$, where $k$ is the wave number, $v$ the wave speed,
$\vec{n}$ the direction of propagation, $\vec{m}$ the normal
to the surface, and $q_\alpha$ an attenuation coefficient.
For a general strain energy function, the $q_{\alpha}$ are the roots of
a bi-cubic equation with positive imaginary parts.
Their explicit analytical expressions are awkward and the method of
Taziev proves useful (Section 4), because it does not require such
expressions.
However, for a whole class of hyperelastic incompressible materials,
inclusive of the Mooney-Rivlin model, the coefficients of attenuation
$q_\alpha$ are obtained analytically, for one is always equal to
$\imi=\sqrt{-1}$, and so the two others are roots of a bi-quadratic
(Section 3).
This property has been touched upon by Flavin but has been only
recently examined in detail (Pichugin \cite{Pich01}).
Here, it leads to the derivation of an \textit{exact and explicit}
secular equation for surface waves, which possesses no more than one
root for the speed.
In the general case, the method of Taziev leads to a
\textit{rationalized} secular equation (a polynomial of degree 12 in
the squared wave speed), with spurious roots to be discarded.

Before these two main results are developed, we summarize in Section 2
the basic governing equations and present the equations of motion as a
first-order linear differential system. The derivation of this system
is a lengthy process and is not obvious at all. 
Thanks to some shorthand notations and to the Stroh formalism, 
the system can however be presented in quite a compact form. 
Its resolution, coupled to the appropriate boundary conditions for 
surface waves (vanishing of the wave away from the interface; 
no incremental traction on the interface), leads to an implicit secular
equation, to be made explicit in the subsequent two Sections.
Section 3 is devoted to the special class of incompressible 
materials which is associated with a factorized propagation condition.
Numerical examples in this section involve a Mooney-Rivlin material
characterized by the same material parameters, state of tri-axial
pre-strain, and normal load as the one considered by Rogerson and
Sandiford \cite{RoSa99}.
A connection is made with their numerical results for the surface wave
speed versus the angle of propagation.
Some new features are highlighted;
in addition, the attenuation coefficients for the partial
displacements are presented.
Section 4 covers the derivation of the secular equation for the
general strain energy density of an incompressible material, 
using a variant of Taziev's technique.
To illustrate the method, we  investigate numerically the case of a
deformed Varga material where a surface wave
propagates in any direction in the plane of shear, with a view to the
non-destructive acoustic evaluation of a deformed rubber insulator.

\section{Preliminaries}
\subsection{Equations of motion}

Consider a half-space, composed of a homogeneous pre-stressed
hyperelastic incompressible material with mass density $\rho$,
characterized by strain energy function $W$.
Let $(O, x_1, x_2, x_3)$ be a fixed rectangular Cartesian coordinate
system such that the body occupies the region $x_2 \ge 0$ and that the
principal stretches coincide with the $\mathcal{O}x_i$ directions, with
corresponding stretch ratios $\lambda_1$,  $\lambda_2$,  $\lambda_3$
($\lambda_1 \ne \lambda_2 \ne \lambda_3 \ne \lambda_1$ and
$\lambda_1 \lambda_2 \lambda_3 =1$).
The half-space is maintained in the static state of
deformation by the application of the constant tractions $\sigma_1$,
$\sigma_2$, $\sigma_3$ at infinity, given by
\begin{equation} \label{p}
  \sigma_i = \lambda_i W_i -p,
\end{equation}
(no summation assumed here)
where $W_i := \partial W/\partial\lambda_i$ and
$p$ is a constant scalar, introduced by the constraint of
incompressibility.

Then consider the superposition of a small-amplitude motion
$\vec{u}(x_1,x_2,x_3,t)$ upon the primary large deformation.
The corresponding incremental nominal stress $\vec{s}$ has the
following components \cite{Ogde84},
\begin{equation} \label{s}
s_{ij} = B_{ijkl}u_{l,k} + pu_{i,j} - p^*\delta_{ij},
\end{equation}
where $p^*$ is a Lagrange multiplier, corresponding to an
increment in $p$, and the non-zero components of the fourth-order
elasticity tensor $\vec{B}$ are (no summation on repeated $i,j$
indexes assumed here)
\begin{align}
& B_{iijj} =
\lambda_i \lambda_j W_{ij},
\nonumber \\
& B_{ijij} =
  (\lambda_i W_i-\lambda_j W_j)\lambda_i^2/(\lambda_i^2-\lambda_j^2),
\nonumber \\
& B_{ijji}= B_{jiij}=
   B_{ijij} - \lambda_i W_i.
\end{align}
The linearized incremental equations of motion and incompressibility
condition read
\begin{equation} \label{motion1}
s_{ij,i} = \rho u_{j,tt},
\quad
u_{j,j} = 0.
\end{equation}

We specialize our analysis to the consideration of a surface
(Rayleigh) wave, propagating in a principal plane but not in a
principal direction, see Figure 1.
Specifically, it is assumed that the wave is traveling with phase
velocity $v$ and wave number $k$ over
the surface $x_2=0$ in a direction which makes an angle $\theta$ with
$\mathcal{O}x_1$; it decays exponentially away from the boundary $x_2=0$, and
produces no incremental traction at the boundary.

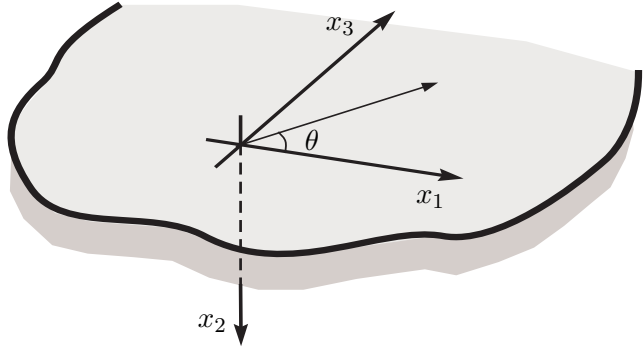
\begin{figure}[!t]
  \setlength{\unitlength}{1mm}
  \begin{center}
  \begin{picture}(88,43)
    \put(0,-7){\resizebox*{88mm}{48mm}{\includegraphics{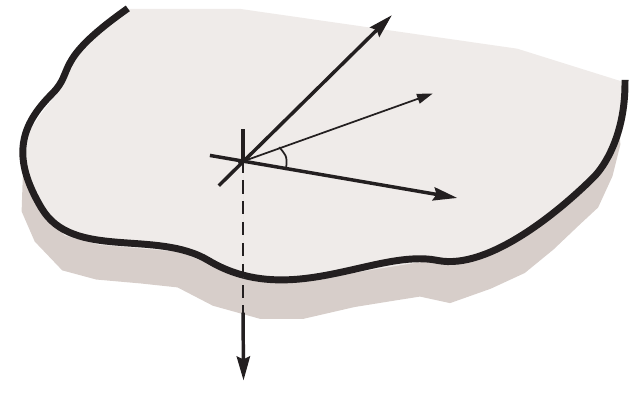}}}
    {\small
      \put(57,13.5){$x_1$} \put(28,-3){$x_2$} \put(45,36.5){$x_3$}
      \put(42.2,20.65){$\theta$}
      }
  \end{picture}
  \end{center}
  \caption{A surface wave propagating in a principal plane but not in a
    principal direction.}
\end{figure}

Hence we model this motion by
\begin{equation} \label{wave}
\{ \vec{u}, p^*, \vec{s} \}
  = \{ \vec{U}(kx_2), \imi kP(kx_2),
          \imi k\vec{S}(kx_2)\}
                e^{\imi k(c_\theta x_1 + s_\theta x_3 - vt)},
\end{equation}
where $c_\theta := \cos\theta$, $s_\theta := \sin\theta$,
and $\vec{U}$, $P$, $\vec{S}$ are functions of $kx_2$ alone.
By substituting these forms of the mechanical displacements and the
tractions into the incremental constitutive equation \eqref{s}
and using \eqref{p}, the incremental equations of motion and the
incompressibility constraint \eqref{motion1} can be cast as a
homogeneous linear system of six first-order differential equations,
\begin{equation} \label{1stOrder}
  \vec{\xi}'=\imi\vec{N}\vec{\xi}\,,
    \quad \text{where} \quad
  \vec{\xi}(kx_2) :=
    [U_1, U_2, U_3, S_{21}, S_{22}, S_{23}]^\text{T},
\end{equation}
within which the prime denotes differentiation with respect to the
variable $kx_2$.

Here the $6\times 6$ real matrix $\vec{N}$ follows the usual block
decomposition \cite{Ting96} of linear anisotropic elasticity,
\begin{equation} \label{N}
\vec{N}  =  \begin{bmatrix}
                    \vec{N}_1 & \vec{N}_2 \\
   \vec{N}_3 + X \vec{1}   & \vec{N}_1^{\text{T}}
                             \end{bmatrix},
 \quad X := \rho v^2,
\end{equation}
where $\vec{N}_1$, $\vec{N}_2 \equiv \vec{N}_2^\text{T}$,
and $\vec{N}_3 \equiv \vec{N}_3^\text{T}$ are $3 \times 3$ matrices.
However, the components of the $\vec{N}_i$ are specific to the
theory of small motions superposed on large static deformations
\cite{Chad97}.
To present them in a compact form, we introduce the short-hand
notations
\begin{align} \label{gammaBeta}
& \gamma_{ij} :=
  (\lambda_i W_i-\lambda_j W_j)\lambda_i^2/(\lambda_i^2-\lambda_j^2)
  \equiv \gamma_{ji} + \lambda_i W_i - \lambda_j W_j,
\nonumber \\
& 2 \beta_{ij} := \lambda_i^2 W_{ii}
   - 2 \lambda_i \lambda_j W_{ij} + \lambda_j^2 W_{jj}
 +  2(\lambda_i W_j-\lambda_j W_i)
         \lambda_i \lambda_j/(\lambda_i^2-\lambda_j^2)
                \equiv 2 \beta_{ji}.
\end{align}
Then $-\vec{N}_1$ and $\vec{N}_2$ are given by
\begin{equation}  \label{N1N2}
 \begin{bmatrix}
       0 & c_\theta (\gamma_{21}-\sigma_2)/\gamma_{21} & 0 \\
       c_\theta & 0 & s_\theta \\
       0 & s_\theta (\gamma_{23}-\sigma_2)/\gamma_{23} & 0
       \end{bmatrix},
\quad
 \begin{bmatrix}
       1/\gamma_{21} & 0 & 0 \\
           0 & 0 & 0 \\
        0 & 0 &  1/\gamma_{23}
      \end{bmatrix},
\end{equation} respectively, and
\begin{equation} \label{N3}
 - \vec{N}_3 = \begin{bmatrix}
                       \eta   &     0     & -\kappa \\
                          0   &     \nu   &      0     \\
                    - \kappa  &     0     &  \mu
                   \end{bmatrix},
\end{equation}
where
\begin{align}
& \eta :=
    2 c_\theta^2(\beta_{12} + \gamma_{21} - \sigma_2)
          + s_\theta^2 \gamma_{31},
\nonumber \\
& \nu :=
  c_\theta^2[\gamma_{12}
      - (\gamma_{21}-\sigma_2)^2 / \gamma_{21}]
   + s_\theta^2[\gamma_{32}
            - (\gamma_{23}-\sigma_2)^2/\gamma_{23}],
\nonumber \\
& \mu :=
    c_\theta^2 \gamma_{13}
       +  2 s_\theta^2 (\beta_{23} + \gamma_{23} - \sigma_2),
\nonumber \\
& \kappa
    :=   c_\theta s_\theta (\beta_{13} - \beta_{12} - \beta _{23}
                             - \gamma_{21} - \gamma_{23} + 2\sigma_2),
\end{align}
(see Destrade \cite{Dest04} for the case
$\sigma_2=0$, and Destrade and Ogden \cite{DeOg04} for the case of
a wave polarized in the symmetry plane of a stretched and sheared
material).

\subsection{Propagation condition}

Now the requirement of exponential decay away from the surface $x_2=0$
is expressed by choosing the following form for the wave,
\begin{equation} \label{u0tau0}
\vec{\xi}(kx_2)
 = \vec{\xi^o} e^{\imi kqx_2},
\quad
\Im(q) >0,
\end{equation}
where $\vec{\xi^o}$ is a constant vector and $q$ is an attenuation
coefficient.
Then the equations of motion \eqref{1stOrder} become the eigenvalue
problem: $\vec{N}\vec{\xi^o} = q \vec{\xi^o}$.
The associated characteristic equation
$\text{det} (\vec{N} - q\vec{1}) =0$, is the
\textit{propagation condition}.
This equation is a cubic in $q^2$, as demonstrated by
Rogerson and Sandiford \cite{RoSa99}:
\begin{equation} \label{bicubic}
\gamma_{21} \gamma_{23} q^6
 - [(\gamma_{21} + \gamma_{23})X - c_1]\,q^4
  + (X^2 - c_2X + c_3)\,q^2
    + (X-c_4)(X-c_5) = 0,
\end{equation}
with
\begin{align}
& c_1 :=  (\gamma_{21}\gamma_{13} + 2\beta_{12}\gamma_{23})c_\theta^2
        + (\gamma_{23}\gamma_{31} + 2\beta_{23}\gamma_{21})s_\theta^2,
\nonumber \\
& c_2 :=  (\gamma_{23} + \gamma_{13} + 2\beta_{12})c_\theta^2
              + (\gamma_{21} + \gamma_{31} + 2\beta_{23})s_\theta^2,
\nonumber \\
& c_3 :=  (\gamma_{12}\gamma_{23} + 2\beta_{12}\gamma_{13})c_\theta^4
        + (\gamma_{21}\gamma_{32} + 2\beta_{23}\gamma_{31})s_\theta^4
\nonumber \\
& \phantom{123456}  + [\gamma_{12}\gamma_{21} + \gamma_{13}\gamma_{31}
                 +\gamma_{23}\gamma_{32}
                   - (\beta_{13} - \beta_{12} - \beta_{23})^2
                    + 4 \beta_{12}\beta_{23}]c_\theta^2 s_\theta^2,
\nonumber \\
& c_4 := \gamma_{12}c_\theta^2 + \gamma_{32}s_\theta^2,
\nonumber \\
& c_5 :=  \gamma_{13}c_\theta^4 + 2 \beta_{13}c_\theta^2s_\theta^2
                   + \gamma_{31}s_\theta^4.
\end{align}
Note that the roots $q_1^2$, $q_2^2$, $q_3^2$ of the bicubic are such
that
\begin{align} \label{Omegas}
& q_1^2 + q_2^2 + q_3^2
   = [(\gamma_{21} + \gamma_{23})X - c_1]/(\gamma_{21} \gamma_{23}),
\nonumber \\
& q_1^2 q_2^2 + q_2^2 q_3^2 + q_3^2 q_1^2
  = (X^2 - c_2X + c_3)/(\gamma_{21} \gamma_{23}),
\nonumber \\
& q_1^2 q_2^2 q_3^2
    = -(X-c_4)(X-c_5)/(\gamma_{21} \gamma_{23}).
\end{align}

\subsection{Implicit secular equation for surface waves}

For the three roots $q_1$, $q_2$, $q_3$ of the propagation
condition \eqref{bicubic} with positive imaginary parts,
the corresponding eigenvalue problems yield three linearly
independent eigenvectors $\vec{\xi}^1$,
$\vec{\xi}^2$,  $\vec{\xi}^3$, respectively.
Then the general solution to the equations of motion \eqref{1stOrder}
may be written as
\begin{equation} \label{xi}
 \mbox{\boldmath $\xi$}(kx_2) =
  \gamma_1 e^{\imi q_1 kx_2}\vec{\xi}^1
   + \gamma_2 e^{\imi q_2 kx_2}\vec{\xi}^2
    +  \gamma_3 e^{\imi q_3 kx_2}\vec{\xi}^3,
\end{equation}
for some constants $\gamma_1$, $\gamma_2$, $\gamma_3$.
Explicitly, $\vec{\xi}^i$ are given by columns
of the matrix adjoint to $\vec{N} - q_i\vec{1}$.
Taking, for example, the second such column gives
$\vec{\xi}^i$ in the form:
\begin{equation}
  \vec{\xi}^i =
 \begin{bmatrix}
   a_4 q_i^4 + a_2 q_i^2 + a_0 \\
   -q_i^5 + b_3 q_i^3 + b_1 q_i \\
  d_4 q_i^4 + d_2 q_i^2 + d_0 \\
  h_3 q_i^3 + h_1 q_i \\
  (\nu - X)(q_i^4 + mq_i^2 + n) \\
  g_3 q_i^3 + g_1 q_i
 \end{bmatrix},
\end{equation}
where expressions for the constants $a_i$, $b_i$, $d_i$, $h_i$,
and $g_i$ are too lengthy to be reproduced here and
\begin{align}
\vspace*{1pt}
&  m = \left(\frac{1}{\gamma_{21}}  +
            \frac{(\gamma_{21} - \sigma_2)^2}
             {\gamma_{21}^2(\nu -X)}c_\theta^2\right) [\eta-X]
 + \left(\frac{1}{\gamma_{23}}
  +\frac{(\gamma_{23}-\sigma_2)^2}{\gamma_{23}^2(\nu -X)}
                       s_\theta^2\right)[\mu-X]
\nonumber \\
& \phantom{1234567890123456}
- 2\kappa\frac{(\gamma_{21}-\sigma_2)(\gamma_{23}-\sigma_2)}
                                    {\gamma_{21}\gamma_{23}(\nu -X)}
                                             c_\theta s_\theta,
\nonumber \\
&  n = \left\{ 1+
  \left[\frac{(\gamma_{21}-\sigma_2)^2}{\gamma_{21}}c_\theta^2
       +\frac{(\gamma_{23}-\sigma_2)^2}{\gamma_{23}}s_\theta^2\right]
                                                    (\nu -X)^{-1}
  \right\}
\nonumber \\
&\phantom{1234567890123456}
    \times [(\mu -X)(\eta-X)-\kappa^2]/(\gamma_{21}\gamma_{23}).
\end{align}
The boundary condition of zero incremental tractions at the plane
surface $x_2=0$ means that
\begin{equation}
  \vec{\xi}(0) =
 [U_1(0), U_2(0), U_3(0), 0, 0, 0]^\text{T}.
\end{equation}
By comparing this expression with \eqref{xi} evaluated for 
$x_2=0$, we conclude that $\gamma_1$, $\gamma_2$, $\gamma_3$
are solutions to a homogeneous linear system of three equations.
The corresponding secular equation is given by
\begin{equation}
 (\nu - X)\begin{vmatrix}
  h_3 q_1^3 + h_1 q_1 & h_3 q_2^3 + h_1 q_2 & h_3 q_3^3 + h_1 q_3
 \\
 q_1^4 + mq_1^2 + n &  q_2^4 + mq_2^2 + n & q_3^4 + mq_3^2 + n \\
 g_3 q_1^3 + g_1 q_1 & g_3 q_2^3 + g_1 q_2 & g_3 q_3^3 + g_1 q_3
 \end{vmatrix} = 0.
\end{equation}
As noted by Taziev \cite{Tazi87} in the context of linear anisotropic
elasticity, this determinant factorizes greatly.
By omitting the factors $\nu - X$, $q_i - q_j$ and $h_1g_3 - h_3g_1$,
we are left with
\begin{equation} \label{secularImplicit}
n \omega_\text{I} = \omega_\text{III}(m-\omega_\text{II}),
\end{equation}
where
\begin{equation} \label{omegas}
\omega_\text{I} := -(q_1 + q_2 + q_3),
\quad
\omega_\text{II} := q_1 q_2 + q_2 q_3 + q_3 q_1,
\quad
\omega_\text{III} := - q_1 q_2 q_3.
\end{equation}
Eq. \eqref{secularImplicit} is the secular equation for surface waves
in deformed incompressible materials; it remains implicit as long as
the explicit expressions for the $\omega_\alpha$ are not known.

\section{Factorization of the propagation condition}

Before moving on to the derivation of an explicit secular
equation, a special case must be treated separately. A simple
analysis shows that, for a certain class of incompressible
hyperelastic materials maintained in a state of large homogeneous
deformation (strain-induced anisotropy), the propagation condition
\eqref{bicubic} factorizes into the product of a term linear in
$q^2$ and a term quadratic in $q^2$, thus leading to simple
explicit expressions for the $q_\alpha$ and eventually, for the secular
equation. This class of materials includes the well-known
Mooney-Rivlin model, often used to describe finite deformations of
rubber. The described factorization does not in general occur for
linear elastic materials such as crystals (intrinsic anisotropy).

\subsection{Conditions on the strain energy function
and explicit secular equation for surface waves}

Following Pichugin \cite{Pich01}, we seek a solution to the
propagation condition \eqref{bicubic} of the form $q^2 = C$,
where $C$ is a constant independent of $v$.
By substituting $q^2 = C$ into \eqref{Omegas},
we obtain three equations for the two quantities $S$
and $P$, the respective sum and product of the remaining $q_\alpha^2$,
namely,
\begin{align}
& S +C = [(\gamma_{21}+\gamma_{23})X-c_1]/(\gamma_{21}\gamma_{23}),
\nonumber \\
& CS + P = (X^2 - c_2X + c_3)/(\gamma_{21}\gamma_{23}),
\label{eqSP}\\
& CP = -(X - c_4)(X - c_5)/(\gamma_{21}\gamma_{23}).
\nonumber
\end{align}
After solving \eqref{eqSP}$_1$ for $S$ and then \eqref{eqSP}$_3$ for
$P$, the substitutions of $S$ and $P$ into \eqref{eqSP}$_2$ allow for
the identification of like-powers of $X$ and $c_\theta^2$ on both
sides of the resulting equation.
Since the identity must hold for all $X$ and $\theta$ we obtain that
$C=-1$, provided the following relationships are satisfied
\begin{equation} \label{conditionFactor}
\gamma_{ij} + \gamma_{ji} = 2 \beta_{ij}.
\end{equation}
Then the associated propagation condition factorizes into
\begin{equation} \label{biquadratic}
(q^2+1)(q^4 - Sq^2 + P) = 0,
\end{equation}
where
\begin{align}
& S = \left(\frac{1}{\gamma_{21}} + \frac{1}{\gamma_{23}}\right)X
       -\left(\frac{\gamma_{12}}{\gamma_{21}}
            + \frac{\gamma_{13}}{\gamma_{23}}\right)c_\theta^2
       -\left(\frac{\gamma_{31}}{\gamma_{21}}
            + \frac{\gamma_{32}}{\gamma_{23}}\right)s_\theta^2,
\nonumber \\
& P = (X - \gamma_{12}c_\theta^2 - \gamma_{32}s_\theta^2)
       (X - \gamma_{13}c_\theta^2 - \gamma_{31}s_\theta^2)
                                       /(\gamma_{21}\gamma_{23}).
\end{align}

To sum up, if the conditions \eqref{conditionFactor} on the strain
energy function are satisfied, then
the bi-cubic \eqref{bicubic} factorizes into the product of a term
linear in $q^2$ and a term quadratic in $q^2$.
It follows that the roots $q_\alpha$ with positive imaginary parts and
hence the corresponding $\omega_\text{I}$, $\omega_\text{II}$,
and $\omega_\text{III}$ defined in \eqref{omegas}, can be found
explicitly: $q_1 = \imi$, $q_2 q_3 = -\sqrt{P}$,
$q_2 + q_3 = \imi\sqrt{2\sqrt{P}-S}$, so that
\begin{equation}
\omega_\text{I} = -\imi\, (1+\sqrt{2\sqrt{P}-S}),
\quad
-\omega_\text{II} = \sqrt{P} + \sqrt{2\sqrt{P}-S},
\quad
\omega_\text{III} = \imi \sqrt{P}.
\end{equation}
The end result is that equation \eqref{secularImplicit} is now an
\textit{explicit secular equation for surface waves in deformed
incompressible materials satisfying \eqref{conditionFactor}},
namely
\begin{equation} \label{explicitSecular}
 n\Bigl(1 + \sqrt{2\sqrt{P}-S}\Bigr)
   + \sqrt{P}\Bigl(m + \sqrt{P} +\sqrt{2\sqrt{P}-S}\Bigr)
    = 0.
\end{equation}

Note that the conditions \eqref{conditionFactor} impose restrictions
upon the strain energy function of the incompressible material.
Explicitly, they read ($i \ne j$)
\begin{equation}
(\lambda_i^2 + 3\lambda_j^2)\lambda_i W_i
 - (3\lambda_i^2 + \lambda_j^2)\lambda_j W_j
  - (\lambda_i^2 - \lambda_j^2)(\lambda_i^2 W_{ii}
                             - 2\lambda_i \lambda_j W_{ij}
                                 + \lambda_j^2 W_{jj}) = 0.
\end{equation}
For example, the Mooney-Rivlin strain energy function,
\begin{equation} \label{Mooney}
W = \mathcal{D}_1(\lambda_1^2 + \lambda_2^2 + \lambda_3^2-3)/2
 +\mathcal{D}_2(\lambda_1^2 \lambda_2^2 + \lambda_2^2 \lambda_3^2
                     + \lambda_3^2 \lambda_1^2-3)/2,
\end{equation}
where $\mathcal{D}_1$ and $\mathcal{D}_2$ are constants, satisfies
this condition. This case is treated in the following subsection.

Note also that  Pichugin \cite{Pich01} finds conditions under which
the propagation condition \eqref{bicubic} admits roots of the form
$q^2 = CX + D$, where $C$, $D$ are constants.
It turns out this possibility arises when the half-space is subject
to an \textit{equi-biaxial} pre-deformation, whatever the strain
energy function may be.
Another simplification occurs when the bi-quadratic in
\eqref{biquadratic} admits a double root (then $S^2 = 4P$);
such is the case for the neo-Hookean form of strain energy function
($\mathcal{D}_2=0$ in \eqref{Mooney}).
The secular equations for surface waves in bi-axially deformed
generic incompressible materials and in tri-axially deformed
 neo-Hookean materials were established by Willson \cite{Will73}
and Flavin \cite{Flav61}, respectively, and are not investigated
further here.

\subsection{Example: Mooney-Rivlin materials}

Now the case of Mooney-Rivlin materials is dealt with in such a way
that a connection is made with the numerical results of
Rogerson and Sandiford \cite{RoSa99}.
For the Mooney-Rivlin strain energy function \eqref{Mooney} the
quantities $\gamma_{ij}$, $\beta_{ij}$, defined in \eqref{gammaBeta},
yield the following forms,
\begin{equation}
\gamma_{ij}
  = (\mathcal{D}_1 + \mathcal{D}_2 \lambda_k^2)\lambda_i^2,
\quad
      2\beta_{ij} =
 (\mathcal{D}_1 + \mathcal{D}_2 \lambda_k^2)(\lambda_i^2+\lambda_j^2)
 = \gamma_{ij} + \gamma_{ji},
\end{equation}
where $k \ne i,j$.

Using these expressions, together with the material parameters
\cite{RoSa99} $\mathcal{D}_{1}=2$, $\mathcal{D}_{2}=0.8$,
the stretch ratios given by $\lambda_{1}^{2}=3.695$,
$\lambda_{2}^{2}=0.7$, $\lambda_{3}^{2}=0.387$, and the
normal load $\sigma_{2}=0.8$, the secular equation
\eqref{explicitSecular} is solved numerically to give the variation
of the surface wave speed with $\theta$, and the graph of Rogerson and
Sandiford \cite{RoSa99} is reproduced with little effort.
We take this opportunity to comment on their statement that
``the surface wave degenerates into a shear wave as $\theta$
approaches 0 and $\pi/2$''.

\begin{figure}[!t]
  \setlength{\unitlength}{1mm}
  \begin{center}
  \begin{picture}(95,70)
    \put(0,0){\resizebox*{95mm}{70mm}{\includegraphics{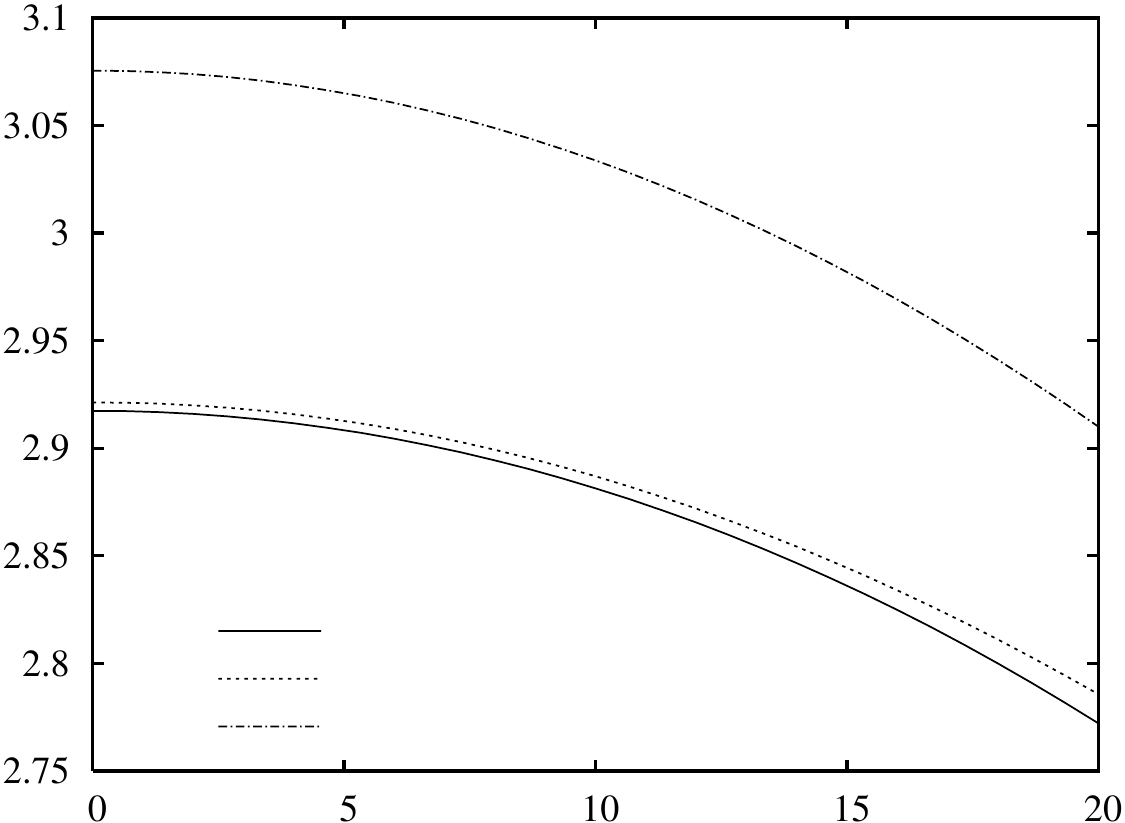}}}
    {\small
      \put(49,-5){$\theta,^{\circ}$} \put(-10,36){$\sqrt\rho v$}
      \put(10.5,15){$\sqrt{X}$}
      \put(10.5,11.5){$\sqrt{c_4}$}
      \put(10.5,7.3){$\sqrt{c_5}$}
      }
  \end{picture}
  \end{center}
  \caption{Scaled speeds of bulk and surface waves near the direction
of greatest stretch in a Mooney-Rivlin material.}
\end{figure}
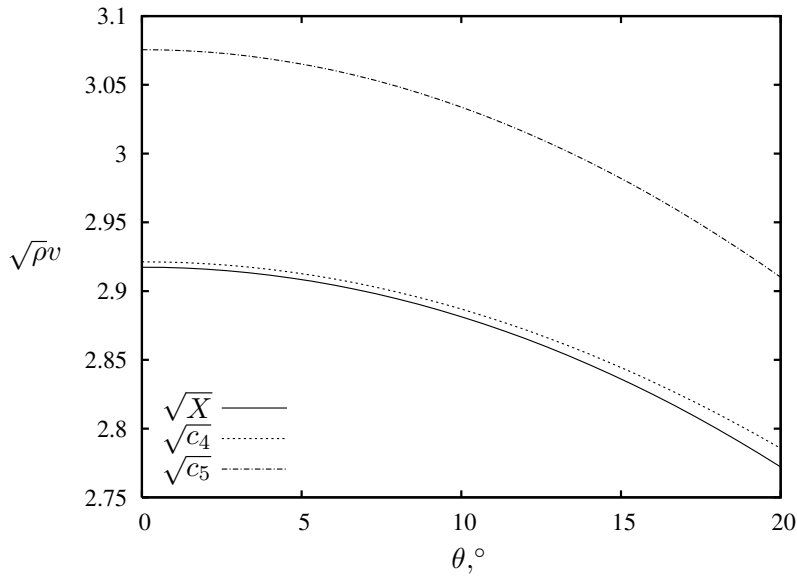

Firstly, we find that at, and close to, the direction $\mathcal{O}x_1$
(also the direction of greatest stretch),
the surface wave speed $v = \sqrt{X/\rho}$ is \textit{distinct} from
the bulk shear wave $\sqrt{c_4/\rho}$, i.e. $c_5 > c_4$ in the
neighborhood of $\theta =0$.
For instance, at $\theta = 0$ the \textit{principal} surface wave
propagation speed $\sqrt{X_0/\rho}$, where $\sqrt{X_0} = 2.917$,
is found from Dowaikh and Ogden's \cite{DoOg90} formula
$X_0 = \gamma_{12} -  \gamma_{21}\zeta^2$, where
\begin{equation} \label{principal0}
  \zeta^3 + \zeta^2
  + \dfrac{2(\beta_{12} + \gamma_{21} -\sigma_2) - \gamma_{12}}{\gamma_{21}}\,\zeta
   - \frac{(\gamma_{21}-\sigma_2)^2}{\gamma_{21}} =0\,,
\end{equation}
(or equivalently, is found from \eqref{explicitSecular}),
while the bulk shear waves propagate at speeds  $\sqrt{c_5/\rho}$
and $\sqrt{c_4/\rho}$ where $\sqrt{c_5} = \sqrt{\gamma_{13}} = 3.076$
and $\sqrt{c_5} = \sqrt{\gamma_{12}} = 2.921$.
Figure 2 shows the variations of these speeds in the
($0^\circ - 20^\circ$) range.
The top (dashed) curve is the graph of $\sqrt{c_5}$,
the middle (dotted) curve is the graph of $\sqrt{c_4}$,
and the bottom (solid) curve is the graph of $\sqrt{X}$.

Secondly, we find that, as the direction of propagation
approaches the $\mathcal{O}x_3$ direction (direction of least stretch,
$\theta =  90^\circ$ here), the surface wave speed
$v = \sqrt{X/\rho}$ is indeed tending to the bulk shear wave speed
$\sqrt{c_5/\rho}$, so that the corresponding
graphs are indistinguishable one from another in the approximative
range ($82^\circ - 90^\circ$).
However, at $\theta =  90^\circ$ exactly, there exists a two-partial
principal surface wave whose speed is \textit{intermediate} between
the bulk shear wave speeds $\sqrt{c_5 / \rho}$ and $\sqrt{c_4 / \rho}$.
Numerically, when $\theta =  90^\circ$, $\sqrt{c_5} = 0.995$,
$\sqrt{c_4} = 1.384$, and this two-partial principal surface wave
propagates with speed $\sqrt{X_{90} / \rho}$ say, where
$\sqrt{X_{90}} = 1.327$, the value found from Dowaikh and Ogden's
\cite{DoOg90} formula $X_{90} = \gamma_{32} - \gamma_{23}\zeta^2$, in which
\begin{equation} \label{principal90}
  \zeta^3 + \zeta^2
  + \dfrac{2(\beta_{23} + \gamma_{23} -\sigma_2) - \gamma_{32}}{\gamma_{23}}\,\zeta
   - \frac{(\gamma_{23}-\sigma_2)^2}{\gamma_{23}} =0\,.
\end{equation}
This peculiar situation is also encountered in cubic crystals
with strong anisotropy, such as nickel \cite{Farn78}.
In short, the subsonic two-partial surface wave must be slower than
any in-plane bulk wave (such as the one propagating with speed
$\sqrt{c_5 / \rho}$), but is indifferent to the anti-plane wave
propagating with speed  $\sqrt{c_4 / \rho}$.
This principal two-partial surface wave is singular because it exists
only in the direction $\theta =  90^\circ$, although ``pseudo-surface
waves'' may be found in its neighborhood.

Once $X = \rho v^2$ is known, the attenuation coefficients $q_i$
are computed from \eqref{biquadratic} and the depth profiles follow
naturally from \eqref{xi}.
Figure 3 displays the imaginary part of the $q_i$,
indicative of the penetration depth, as a function of $\theta$.
The horizontal straight top (dashed) line is for $q_1=\imi$.
The two other curves (dotted and solid) are for the
imaginary parts of $q_2$ and $q_3$.
At $\theta=0$, there are only two partial modes, one corresponding to
$q_1=\imi$, the other corresponding to $q_3 = 0.119\imi$.
At $\theta \gtrsim 0$, a third partial mode appears, corresponding to
a $q_2$ of the form $q_2 = \imi \beta_2$, $\beta_2 \lesssim 0.52$.
In the approximate ranges $(0^\circ - 23^\circ)$ and
$(68^\circ - 90^\circ)$, the attenuation factors are of the form
$q_1 = \imi$, $q_2 = \imi \beta_2$, $q_3 = \imi \beta_3$ with
$\beta_2 >0$, $\beta_3>0$.
In the approximate range $(23^\circ - 68^\circ)$,
they are of the form $q_1 = \imi$, $q_2 = \alpha + \imi \beta$,
$q_3 = -\alpha + \imi \beta$ with $\beta >0$.
As $\theta$ approaches $90^\circ$, and $X$ approaches $c_5$,
the imaginary part of the attenuation factor $q_3$ tends to zero,
indicating a deeply penetrating quasi-bulk surface wave \cite{Dari98}.
Finally at $\theta =  90^\circ$, a singular two-partial principal
surface wave exists, with one mode corresponding to $q_1 = \imi$ and
the other corresponding to $q_2 = 0.212 \imi$
(this latter value corresponds to a discontinuity in the
representation of $q_2$ as a function of $\theta$ and cannot be
represented on the graph).

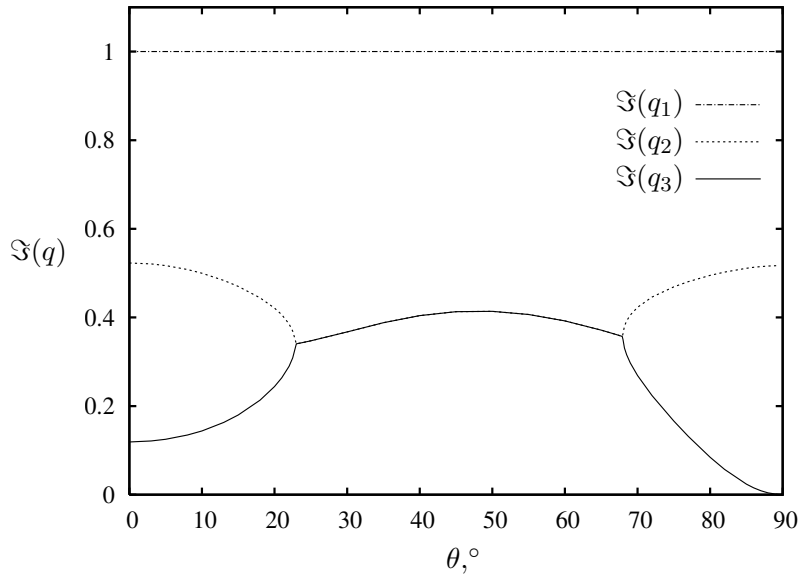
\begin{figure}[!t]
  \setlength{\unitlength}{1mm}
  \begin{center}
  \begin{picture}(95,70)
    \put(0,0){\resizebox*{95mm}{70mm}{\includegraphics{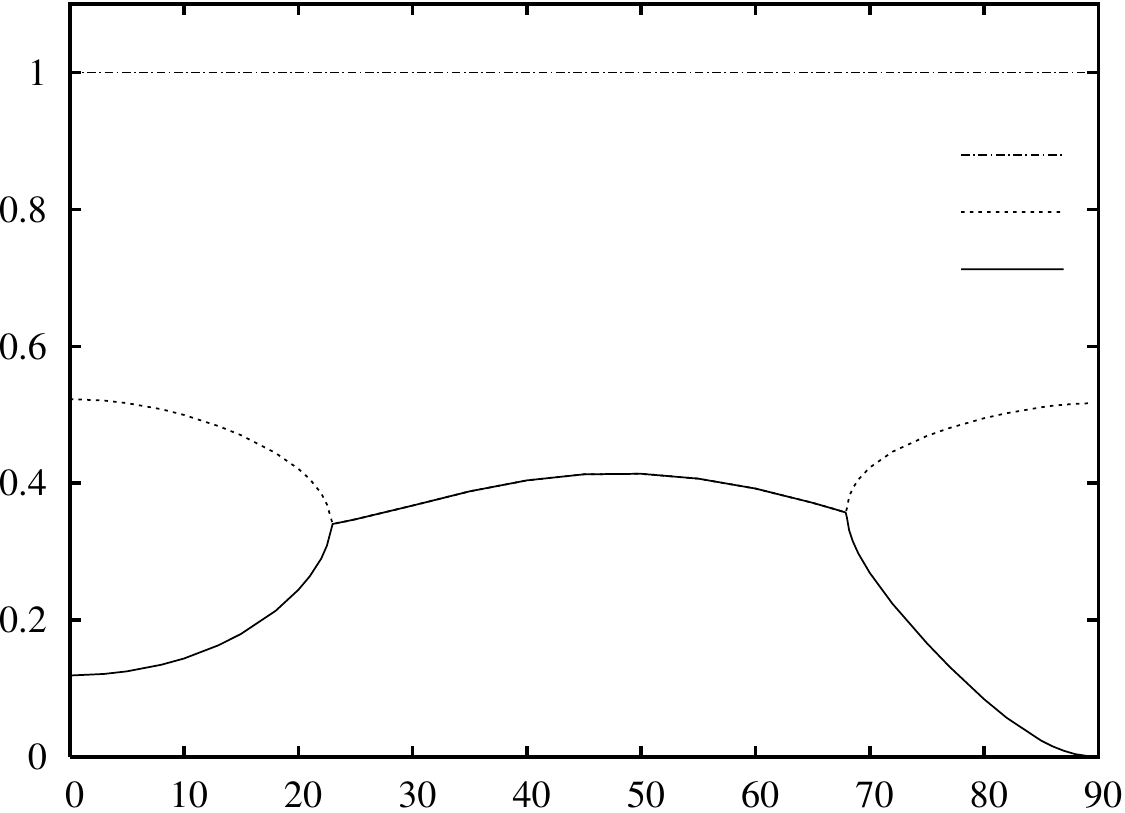}}}
    {\small
      \put(48,-5){$\theta,^{\circ}$} \put(-10,36){$\Im(q)$}
      \put(70.5,55.8){$\Im(q_1)$}
      \put(70.5,50.8){$\Im(q_2)$}
      \put(70.5,45.8){$\Im(q_3)$}
      }
  \end{picture}
  \end{center}
  \caption{Imaginary part of the attenuation coefficients
    for a surface wave in a deformed Mooney-Rivlin material.}
\end{figure}

\section{General case}

\subsection{Explicit secular equation for surface waves}

In the general case, where no factorization of the propagation
condition occurs, a different treatment is required.
The ``method of the polarization vector'', introduced by Currie
\cite{Curr79}, refined by Taziev \cite{Tazi89}, and recently
revisited by Ting \cite{Ting04}, proves to be a most effective mean
for deriving the secular equation as a polynomial in $X=\rho v^{2}$.
It relies on the equations
\begin{equation} \label{UKU}
\overline{\vec{U}}(0) \cdot\vec{K}^{(n)}\vec{U}(0)=0,
\end{equation}
where $\vec{K}^{(n)}$ is the symmetric $3 \times 3$ lower left
submatrix of the $\vec{N}^{n}$ matrix and $n$ is an integer.
Hence $\vec{K}^{(1)} =\vec{N}_{3} + X \vec{1}$,
$\vec{K}^{(2)} = \vec{K}^{(1)}\vec{N}_1
+ \vec{N}_1^\text{T}\vec{K}^{(1)}$, etc.

Also, for a wave propagating in the symmetry plane of a monoclinic
crystal, Ting \cite{Ting04}
showed that $\vec{U}(0)$ is of the form
\begin{equation}
  \vec{U}(0)=U_{1}(0)[1,\imi\alpha_{2}, \beta_{1}]^\text{T},
\end{equation}
where $\alpha_{2}$, $\beta_{1}$ are real numbers.
We checked that $\vec{U}(0)$ is also of this form in the present
case of a surface wave propagating in a principal plane of a deformed
material.

Computing $\vec{N}^{-1}$ and $\vec{N}^{3}$, we find that
$K_{12}^{(n)}=K_{23}^{(n)}=0$, for $n=-1,1,3$.
It follows that the equations \eqref{UKU}
written for $n=-1,1,3$ reduce to the non-homogeneous system
\begin{equation} \label{systemK}
\begin{bmatrix}
K_{13}^{(-1)} & K_{33}^{(-1)} & K_{22}^{(-1)} \\
K_{13}^{(1)} & K_{33}^{(1)} & K_{22}^{(1)} \\
K_{13}^{(3)} & K_{33}^{(3)} & K_{22}^{(3)} \end{bmatrix}
\begin{bmatrix} 2 \beta_{1}\\
\beta_{1}^{2}\\
\alpha_{2}^{2}\end{bmatrix}
=
\begin{bmatrix}-K_{11}^{(-1)}\\
 -K_{11}^{(1)}\\
-K_{11}^{(3)}\end{bmatrix}.
\end{equation}
By Cramer's rule,
we find $2 \beta_{1}=\Delta_{1}/\Delta$,
$\beta_{1}^{2}=\Delta_{2}/\Delta$ where
\begin{multline} \label{def_Delta}
\Delta=
\begin{vmatrix}
K_{13}^{(-1)} & K_{33}^{(-1)} & K_{22}^{(-1)} \\
K_{13}^{(1)} & K_{33}^{(1)} & K_{22}^{(1)} \\
K_{13}^{(3)} & K_{33}^{(3)} & K_{22}^{(3)}
\end{vmatrix},
\\
\Delta_{1}=
\begin{vmatrix}
-K_{11}^{(-1)} & K_{33}^{(-1)} & K_{22}^{(-1)} \\
-K_{11}^{(1)}  & K_{33}^{(1)}  & K_{22}^{(1)}  \\
-K_{11}^{(3)}  & K_{33}^{(3)}  & K_{22}^{(3)}
\end{vmatrix},\qquad\quad\;
\\
\Delta_{2}=
\begin{vmatrix}
K_{13}^{(-1)} & -K_{11}^{(-1)} & K_{22}^{(-1)} \\
K_{13}^{(1)}  & -K_{11}^{(1)}  & K_{22}^{(1)} \\
K_{13}^{(3)}  & -K_{11}^{(3)}  & K_{22}^{(3)}
\end{vmatrix},
\end{multline}
so that
\begin{equation} \label{expl_secular}
\Delta_{1}^{2}-4\Delta\Delta_{2}=0,
\end{equation}
which is the \textit{explicit secular equation for non-principal
surface waves in deformed incompressible materials}.

Upon inspection of \eqref{N} and \eqref{N3}, we find that
$K_{13}^{(1)}=\kappa$, $K_{33}^{(1)}=X-\mu$, $K_{22}^{(1)}=X-\nu$,
$K_{11}^{(1)}=X-\eta$. Computing $\vec{N}^{-1}$, we find that up
to a common disposable factor, $K_{13}^{(-1)}$, $K_{33}^{(-1)}$,
$K_{22}^{(-1)}$, $K_{11}^{(-1)}$ are proportional to polynomials
of degree 1, 2, 3, 2 in $X$, respectively. Similarly, computing
$\vec{N}^{3}$, we find that $K_{13}^{(3)}$, $K_{33}^{(3)}$,
$K_{22}^{(3)}$, $K_{11}^{(3)}$ are polynomials of degree 1, 2, 1,
2, respectively. We conclude from the definitions
\eqref{def_Delta} of $\Delta$, $\Delta_{1}$, $\Delta_{2}$ that the
secular equation \eqref{expl_secular} is a polynomial of degree 12
in $X=\rho v^2$, just as for monoclinic crystals in linear
anisotropic elasticity \cite{Tazi87}. 
It is too long to reproduce here but it was obtained in a 
formal manner with Maple and with Mathematica.

The numerical resolution of the polynomial \eqref{expl_secular}
yields a priori 12 roots for $X$. From these, we discard at once
the complex roots, the negative real roots, and the roots
corresponding to supersonic surface waves (faster than bulk
waves). Out of the remaining roots, at most one will yield
attenuation coefficients $q_1$, $q_2$, $q_3$ (with a positive
imaginary part) from the propagation condition \eqref{bicubic},
such that the exact secular equation \eqref{secularImplicit} is
satisfied.

To conclude this Section, we check that the rationalized secular
equation  \eqref{expl_secular} is consistent with the known secular
equation for principal surface waves.
When $\theta = 0$, it is easy to see
that $K_{13}^{(-1)} = K_{13}^{(1)} = K_{13}^{(3)} = 0$, therefore
$\Delta = \Delta_2 =0$. Thus, \eqref{expl_secular} reduces to $\Delta_1 = 0$.
By substituting  $X = \gamma_{12} - \gamma_{21} \zeta^2$,
we find that $\Delta_1$ factorizes into the product of a quadratic
in $\zeta^2$ and two cubics in $\zeta$, one of which is indeed
Dowaikh and Ogden's \cite{DoOg90} equation~\eqref{principal0}.

\subsection{Example: Varga materials}

The standard Varga strain energy function \cite{Varg66, Hill01} is
defined as
\begin{equation}
W = \mathcal{C}(\lambda_1 + \lambda_2 + \lambda_3 -3),
\end{equation}
where the material parameter $\mathcal{C}$ is constant.
This strain energy function has been introduced to describe natural
rubber vulcanizates.
It leads to the following expressions for the quantities
$\gamma_{ij}$ and $\beta_{ij}$, defined in
\eqref{gammaBeta},
\begin{equation} \label{moduliVarga}
\gamma_{ij}
  = \mathcal{C} \lambda_i^2 / (\lambda_i + \lambda_j),
\quad \beta_{ij} =
 \mathcal{C} \lambda_i \lambda_j
 / (\lambda_i + \lambda_j).
\end{equation}

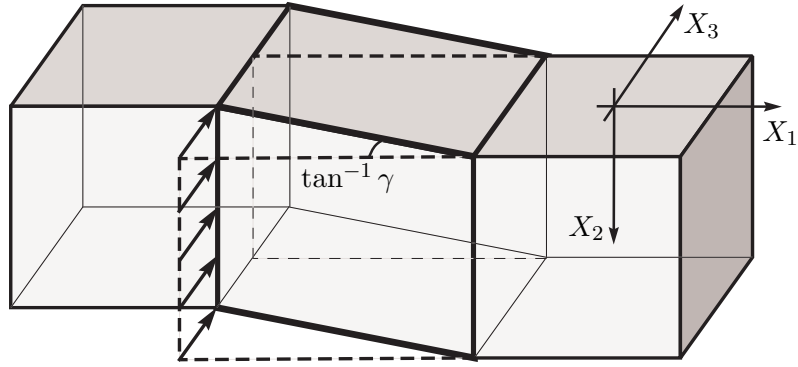
\begin{figure}[!h]
  \setlength{\unitlength}{1mm}
  \begin{center}
  \begin{picture}(105,50)
    \put(0,-2){\resizebox*{105mm}{50mm}{\includegraphics{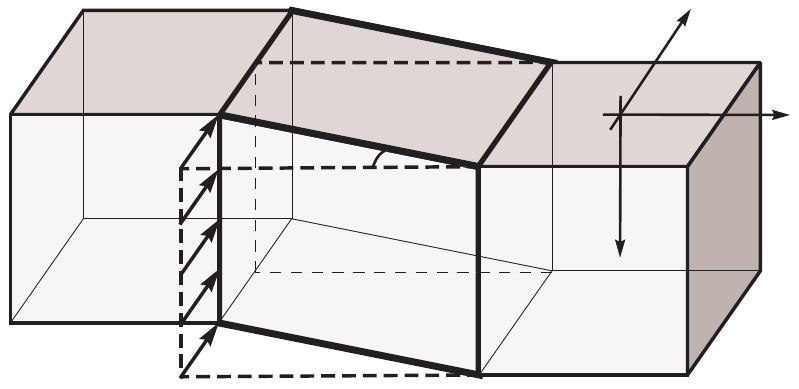}}}
    {\small
      \put(101.3,28.7){$X_1$} \put(75.5,16){$X_2$} \put(90.8,43){$X_3$}
      \put(40,22.5){$\tan^{-1}\gamma$}
      }
  \end{picture}
  \end{center}
  \caption{An elastic material under simple shear;
    the dashed lines represent the body at rest.}
\end{figure}

Simple shear plays an important role in the experimental determination
of a strain energy function \cite{Ogde84}.
Consider now a half-space made of Varga material,
subject to an amount of shear $\gamma$ along $X_3$, see Figure 4.
With a view to the possible non-destructive evaluation of sheared
rubber, we are interested in the propagation of a surface wave in any
direction in the plane of shear $\mathcal{O}X_1X_3$.
We assume that this plane is free of normal load, thus
$\sigma_{2} = 0$.

The principal stretches associated with the amount of shear $\gamma$
may be determined using the following equations \cite{CoOg95},
\begin{equation}
  \lambda_1-\lambda_1^{-1}=\gamma\,,\qquad \lambda_1>1\,,\qquad \lambda_2=1\,,\qquad
  \lambda_3=\lambda_1^{-1}\lambda_2^{-1}\,.
\end{equation}
We substitute the parameters \eqref{moduliVarga} into the secular
equation \eqref{expl_secular}, and then solve that equation
numerically for the amounts of shear: $\gamma = 0.5, 1.0, 1.5$,
corresponding to the angles of shear:
$\tan^{-1}\gamma =  26.56^{\circ}, 45^{\circ}, 56.31^{\circ}$,
respectively.
We obtain the velocity of the surface wave propagating in a direction
making an angle $\theta + \psi$ with $\mathcal{O}X_1$, where
$\psi = (1/2)\tan^{-1}(2/\gamma) = 37.98^{\circ}, 31.72^{\circ},
26.56^{\circ}$, respectively.
Note that here $\mathcal{O}X_1$ is not a principal axis, for the two principal
axes $\mathcal{O}x_1$ and $\mathcal{O}x_3$ in the plane of shear are at an angle $\psi$
with $\mathcal{O}X_1$ and with $\mathcal{O}X_3$, respectively.

For the selection of the relevant root of the secular equation out
of the 12 possible ones, we followed the checking procedure described
at the end of the previous Section.
For example, at $\gamma = 1$ and $\theta + \psi =0$, the rationalized
secular equation  \eqref{expl_secular} has only 2 real roots, giving
$\sqrt{\rho v^2 / \mathcal{C}} = 0.848$ or  $\sqrt{\rho v^2 / \mathcal{C}} = 0.884$.
Both roots yield three attenuation factors with a positive imaginary
part from the propagation condition \eqref{bicubic}.
However, the exact secular equation \eqref{secularImplicit} is
satisfied only with the first root.
Hence, with a 32 digit precision, Maple finds that 
$|n \omega_\text{I} / 
    [\omega_\text{III}(m-\omega_\text{II})] - 1| 
                                      < 10^{-22}$ 
when  $\sqrt{\rho v^2 / \mathcal{C}} = 0.848$, indicating that 
\eqref{secularImplicit} is satisfied; on the other hand, 
$|n \omega_\text{I} / 
    [\omega_\text{III}(m-\omega_\text{II})] - 1| 
                                      > 1.96$ 
when  $\sqrt{\rho v^2 / \mathcal{C}} = 0.884$, indicating that 
\eqref{secularImplicit} is not satisfied then.

\begin{figure}[!t]
  \setlength{\unitlength}{1mm}
  \begin{center}
  \begin{picture}(95,70)
    \put(0,0){\resizebox*{95mm}{70mm}{\includegraphics{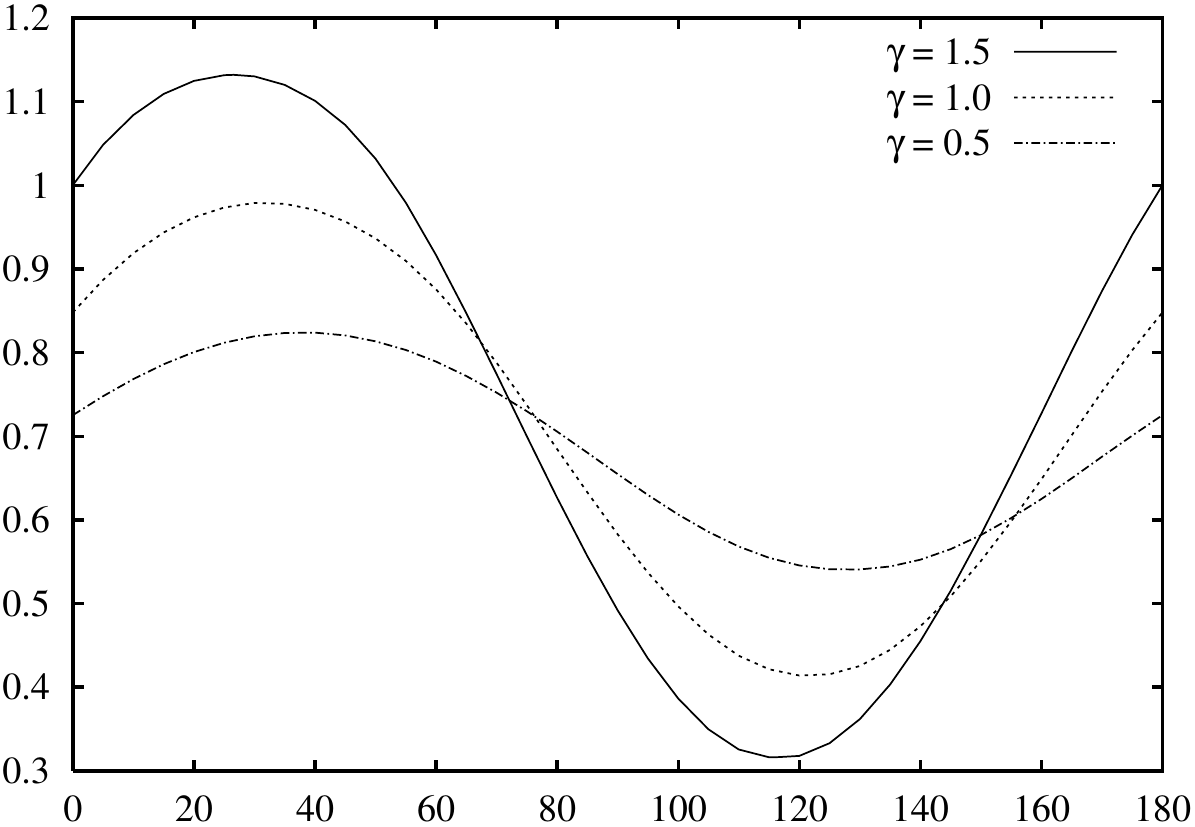}}}
    {\small
      \put(45,-5){$\theta+\psi,^{\circ}$}
      \put(-14,35){${\displaystyle \sqrt{\frac{\rho v^2}{\mathcal{C}}}}$}
     }
  \end{picture}
  \end{center}
  \caption{Scaled surface wave speed as a function of $\theta + \psi$
    for a Varga material subjected to simple shear deformations.}
\end{figure}

Figure 5 displays the dependence of $\sqrt{\rho v^2 /
\mathcal{C}}$ on the angle $\theta+\psi$ over the range $[0^\circ
- 180^\circ]$. The solid, dot, and dash-dot curves correspond to
an amount of shear of 0.5, 1.0, 1.5, respectively. The figure
confirms what is to be expected intuitively: as the half-space is
more and more sheared, the strain-induced anisotropy increases,
and its influence on the surface wave speed is more and more
marked. It also shows that the surface wave travels at its fastest
(slowest) speed along the direction of greatest (least) stretch,
thus allowing for an acoustic determination of the directions of
the principal stretches in sheared rubber.




\end{document}